\documentclass[aps,pra,floatfix,twocolumn,showpacs]{revtex4}
\usepackage{epsfig}
\usepackage{graphicx}
\usepackage{dcolumn}
\begin{document}
\title{Calculation of the energy levels of Ge, Sn, Pb and their ions 
in the $V^{N-4}$ approximation}
\author{V. A. Dzuba}
\email{V.Dzuba@unsw.edu.au}
\affiliation{School of Physics, University of New South Wales, Sydney 2052,
Australia}

\date{\today}

\begin{abstract}
Energy levels of germanium, tin and lead together with their single, double 
and triple ionized positive ions have been calculated using the $V^{N-M}$
approximation suggested in the previous work (Dzuba, physics/0501032)
($M=4$ - number of valence electrons).
Initial Hartree-Fock calculations are done for the quadruply ionized ions
with all valence electrons removed. The core-valence correlations are
included beyond the second-order of the many-body perturbation theory.
Interaction between valence electrons is treated by means of the
configuration interaction technique. It is demonstrated that accurate treatment
of the core-valence correlations lead to systematic improvement of the
accuracy of calculations for all ions and neutral atoms.

\end{abstract}

\maketitle

\section{introduction}

This work further develops the $V^{N-M}$ approximation suggested in Ref.~\cite{vn}.
It also presents the details of the calculations of the energy levels of
Ge~II, Sn~II and Pb~II \cite{Dzuba05} which were needed to study their
dependence on the fine structure constant $\alpha$ ($\alpha = e^2/\hbar c$).
Some lines of Ge~II, Sn~II and Pb~II have been observed in quasar absorption
spectra and the information on the dependence of corresponding frequencies
on $\alpha$ is needed to study possible variation of the fine structure
constant at early epoch.

In the vicinity of the physical value of $\alpha$ the 
frequency of an atomic transition can be presented in a form
\begin{equation}
  \omega = \omega_0 +qx,
\label{qx}
\end{equation}
where $x=(\alpha/\alpha_0)^2-1$ and $\alpha_0$ and $\omega_0$ are the present-day 
laboratory values of the fine structure constant and transition frequency.

The values of the $q$-coefficients can only be found from atomic calculations
by, e.g., varying the value of $\alpha$ in computer codes based on relativistic 
equations. In many cases calculated values of the $q$-coefficients are more 
stable than the energies. This is because they are not sensitive to 
incompleteness of the basis set with respect to the principal quantum number
$n$. Indeed, relativistic corrections are proportional to $1/\nu^3$ 
\cite{Dzuba99} ($\nu$ is the effective principal quantum number) while energies 
are proportional to  $1/\nu^2$. If we include more states of high $\nu$ this
would have greater effect on the energies than on relativistic corrections
presented by $q$-coefficients.

However, in the case of strong configuration mixing and level pseudo-crossing
calculation of $q$-coefficients may become very unstable \cite{Dzuba02}.
In the vicinity of level pseudo-crossing the values of $q$-coefficients change
very rapidly with $\alpha$ and small error in determining the position of the
level crossing may lead to large error in the values of $q$.

Level pseudo-crossing always means strong configuration mixing between the
states. However, strong configuration mixing may also take place 
without level pseudo-crossing. This can also cause instability in
calculated values of $q$-coefficients. Indeed, relativistic correction
to the energy of a single electron state $|njlm\rangle$ strongly
depends on the total momentum $j$ of this state 
(see, e.g. formula~(7) in Ref.~\cite{Dzuba99}).
Therefore configurations composed from states of different $j$ may
have very different values of $q$ and small error in the
the configuration mixing coefficients would lead to large error in
the resulting $q$ value for the mixed state~\cite{comment}.

Strong configuration mixing and level pseudo-crossing 
take place for Ge~II, Sn~II and Pb~II ions \cite{Dzuba05}
as well as for many other atoms and ions \cite{Dzuba02}. This means that
calculations need to be done to very high accuracy to ensure stable values of 
the $q$-coefficients.
The criterion is that deviation of the calculated energies from the
experimental values must be much smaller than the experimental energy
interval between mixed states. 

There are many other areas of research where accurate atomic calculations 
are needed.
These include parity and time invariance violation in atoms 
(see, e.g.~\cite{ginges}), atomic clocks \cite{clock}, interaction
of positrons with atoms~\cite{Gribakin}, etc.

A way to do accurate calculations for atoms with several $s$ and/or 
$p$ valence electrons has been suggested in Ref.~\cite{vn}. 
It is called ``the $V^{N-M}$ approximation'', where
$V$ is the Hartree-Fock potential created by $N-M$ electrons of the
closed shell ion, $N$ is total number of electrons in neutral atom and 
$M$ is the number of valence electrons.
Initial Hartree-Fock calculations are done for a closed-shell positive 
ion with all valence electrons removed. It has been demonstrated in 
Ref.~\cite{vn} that the Hartree-Fock potential of the closed-shell
positive ion is often a good starting approximation for a neutral
atom. This is the case when valence electrons are localized on
distances larger than the size of the core. Then they can affect
only energies of core states but not their wave functions.
Since the potential created by core electrons depends on the
electron charge density and does not depend on electron energies
it doesn't matter which core states are used to calculate the 
potential - states of the neutral atom or states of the closed-shell
positive ion.

The effective Hamiltonian for valence electrons is
constructed using the configuration interaction (CI) technique.
Core-valence correlations are included by adding the
electron correlation operator $\hat \Sigma$ to the CI Hamiltonian. 
Many-body perturbation theory (MBPT) is used to calculate $\hat \Sigma$. 
The main advantage of the $V^{N-M}$ approximation is that MBPT is 
relatively simple (no subtraction diagrams) and the $\hat \Sigma$ 
operator can be calculated beyond the second-order of the MBPT. 
It has been demonstrated in Ref.~\cite{vn} that inclusion of the 
higher-order core valence correlations lead to further significant 
improvement of the accuracy of calculations.

In the previous work~\cite{vn} the $V^{N-M}$ approximation was used 
for Kr and Ba while higher-order core-valence correlations were
included for Ba and Ba$^+$ only. In the present work we study twelve
complicated many-electron systems including germanium, tin, lead
and their positive ions. We demonstrate that using the $V^{N-4}$
approximation ($M=4$ for the case of Ge, Sn and Pb) and
accurate treatment of the core-valence correlations lead to
high accuracy of calculations for all twelve systems.
This indicates that the $V^{N-M}$ approximation is a
good approximation for a wide range of atoms and ions.

\section{Calculations}

The effective Hamiltonian for valence electrons in the $V^{N-M}$ 
approximation has the form
\begin{equation}
  \hat H^{\rm eff} = \sum_{i=1}^M \hat h_{1i} + \sum_{i \neq j}^M \hat h_{2ij} ,
\label{heff}
\end{equation}
$\hat h_1(r_i)$ is the one-electron part of the Hamiltonian
\begin{equation}
  \hat h_1 = c \mathbf{\alpha \cdot p} + (\beta -1)mc^2 - \frac{Ze^2}{r} + V^{N-4}
 + \hat \Sigma_1.
\label{h1}
\end{equation}
$\hat \Sigma_1$ is the correlation potential operator which is exactly the
same in the $V^{N-M}$ approximation as for the single-valence electron atoms
(see, e.g.~\cite{CPM}). It can be calculated in the second-order of the MBPT.
Selected chains of the higher-order diagrams can be included into
$\hat \Sigma_1$ in all orders using technique developed for single-valence
electron atoms (see, e.g.~\cite{Dzuba89}). 

$\hat h_2$ is the two-electron part of the Hamiltonian
\begin{equation}
  \hat h_2 = \frac{e^2}{|\mathbf{r_1 - r_2}|} + \hat \Sigma_2(r_1,r_2),
\label{h2}
\end{equation}
$\hat \Sigma_2$ is the two-electron part of core-valence correlations. 
It represents screening of Coulomb interaction between valence electrons by 
core electrons. We calculate $\hat \Sigma_2$ in the second order of MBPT. 
Inclusion of the higher-order correlations into $\hat \Sigma_2$ will be a 
subject of further study. However, the calculations show that in most  
cases accurate treatment of $\hat \Sigma_1$ is more important than for 
$\hat \Sigma_2$.
The details of the calculation of $\hat \Sigma_1$ and $\hat \Sigma_2$ can be 
found elsewhere \cite{CPM,Dzuba89,Kozlov96, Johnson98}. 
Note however that in contrast to the previous works~\cite{Kozlov96,Johnson98}
we have no so called {\em subtraction diagrams}.

Number of electrons $M$ is the only parameter in the effective Hamiltonian~(\ref{heff}) 
which changes when we move between different ions of the same atom.
The terms $V^{N-4}$, $\hat \Sigma_1$
and $\hat \Sigma_2$ remain exactly the same.

The form of the effective Hamiltonian is also the same for 
all ions if some other potential $V$ is used to generate the core states.
However, the $\hat \Sigma$ operator would have terms proportional to 
$V^{N-4} - V$ (subtraction diagrams \cite{Kozlov96}). In the $V^{N-M}$
approximation $V \equiv V^{n-4}$ and subtraction diagrams disappear.
The MBPT becomes relatively simple which makes it easier to include
higher-order core-valence correlations.

\subsection{Electron shell structure of lead.}

\begin{figure}
\centering
\epsfig{figure=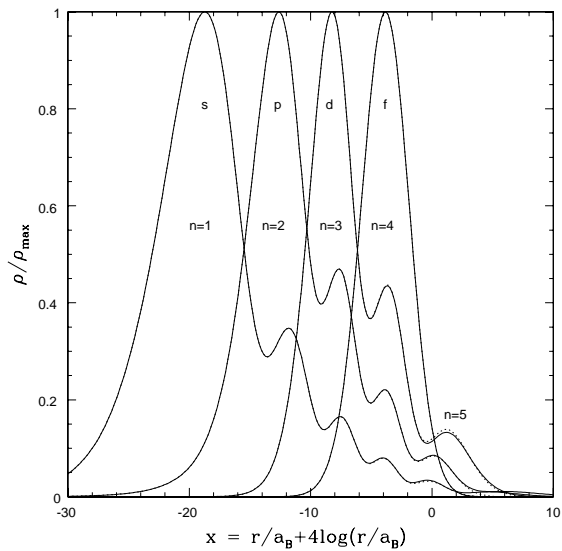,scale=0.38}
\caption{Electron density of the $s,p,d$ and $f$ electrons of Pb~I and Pb~V
as explained in the text.}
\label{pb1}
\end{figure}

\begin{figure}
\centering
\epsfig{figure=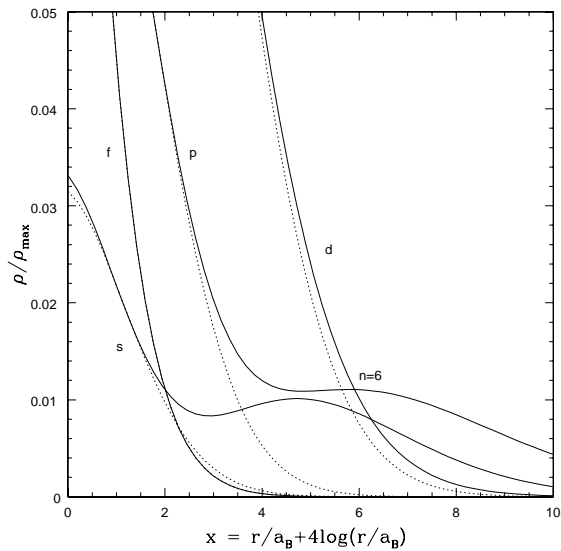,scale=0.38}
\caption{Details of electron densities of Pb~I and Pb~V at large distances.}
\label{pb2}
\end{figure}

To understand how the $V^{N-M}$ approximation works it is very instructive 
to look at electron shells of a many-electron atom. We chose lead because it
is the heaviest of the considered atoms. It probably has the richest possible
electron shell structure.  Neutral lead has eighty two electrons occupying six shells.
Angular momentum $l$ ranges from 0 ($s$-electrons) to 3 ($f$-electrons).
Figs.~\ref{pb1} and \ref{pb2} present electron densities of Pb~I (solid line)
and Pb~V  (dotted line) separately for $s$, $p$, $d$ and $f$ electrons. 
The density is the sum over principal quantum number $n$, total momentum $j$ 
and its projection $m$ while angular momentum $l$ is fixed:
\[    \rho(r)_l= \sum_{njm} |\psi(r)_{njlm}|^2r^2. \]
The values of $\rho(r)_l$ in the maximum are very different for different
$l$. Therefore, we present normalized functions $\rho(r)_l/\rho_{max}$ to be able 
to fit all graphs into one diagram. 

Electron shell structure can be clearly seen on Fig.~\ref{pb1}. Each density
has a local peak at $n-l=1,2,$ etc. The position of the peak depends mostly
on $n$ and is about the same for all $l$. This means that all electrons with
the same $n$ are localized at about the same distances regardless of their
angular momentum $l$, thus making a shell.

The difference between Pb~I and Pb~V cannot be seen on Fig.~\ref{pb1}.
Fig.~\ref{pb2} presents details of the right bottom corner of the Fig.~\ref{pb1}.
Dotted lines which correspond to electron densities of the Pb~V ion show no
peak at $n=6$ because of absence of the $6s$ and $6p$ electrons. The
removal of four valence electrons has some effect on the 
density of $d$-electrons at about the same distances where the $6s$ and 
$6p$ electrons are localized and practically no effect on the densities 
of all electrons on shorter distances. This is because valence electrons 
are localized on large distances and they can only create constant potential 
in the core which can change the energies of the core states but cannot 
change their wave functions.

\begin{figure}
\centering
\epsfig{figure=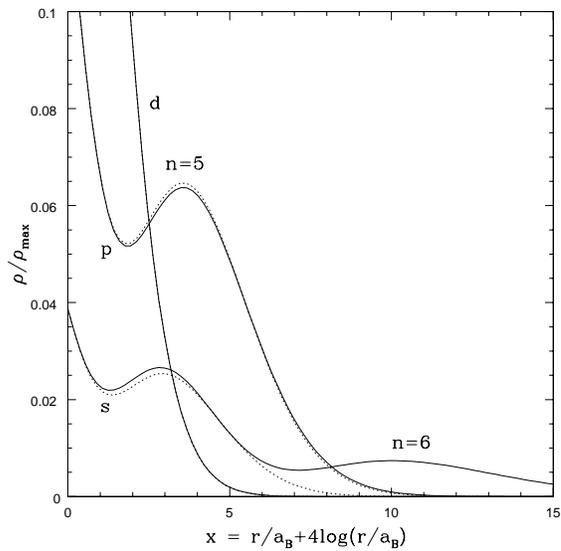,scale=0.38}
\caption{Electron densities of Ba~I and Ba~III at large distances.}
\label{ba}
\end{figure}

One can see from Fig.~\ref{pb2} that there is an overlap between
the wave functions of valence electrons of Pb~I ($6s$ and $6p$ electrons)
and the wave function of the core outermost state $5d$. We have presented
for comparison on Fig.~\ref{ba} the electron densities of Ba~I and Ba~III on
large distances. It is easy to see that the overlap between core and valence
electrons in barium is much smaller than the overlap between core and
valence electrons in lead. As a consequence, outermost core state of barium
($5p$) is much less affected by removal of two $6s$ electrons than 
compared to the effect of removal of two $6s$ and two $6p$ electrons
on the $5d$ state of lead. This means that the $V^{N-2}$ approximation
for Ba should work much better than the $V^{N-4}$ approximation for Pb.
The situation is exactly the same as for the $V^{N-1}$ approximation
for atoms with one external electron. It is very well known that the
$V^{N-1}$ approximation works extremely well for alkali atoms and not
so well for atoms like Ga, In, Tl, etc. The reason is the same in both
cases. Valence electrons must not overlap with the core for the 
$V^{N-M}$ to be good starting approximation regardless of whether
$M=1$ or $M > 1$.

Similar to the fact that the $V^{N-1}$ approximation is a good approximation
for thallium, although not as good as for alkali atoms, the $V^{N-M}$
approximation is a good approximation for Pb, Sn and Ge, although not as
good as for Ba.

Below we present specifics of calculations for germanium, tin and lead.

\subsection{Calculations for germanium}

Germanium is the lightest of three atoms ($Z$=32) and the easiest
from computational point of view. 
Its ground state configuration is $1s^22s^22p^63s^23p^63d^{10}4s^24p^2$.
The core-valence correlations are relatively 
small due to small number of electrons in the core. 

We calculate $\hat \Sigma_1$ and $\hat \Sigma_2$ for the effective Hamiltonian
(\ref{heff}) in the second order of the MBPT. Inclusion of $\hat \Sigma_1$
brings single-electron energies of Ge~IV to agreement with the experiment
on the level of 0.1\%. No higher-order core-valence correlations need to
be included. 

In fact, inclusion of the higher-order correlations using technique developed
in Ref.~\cite{Dzuba89} doesn't lead to better results for germanium. 
This is because the technique was developed for heavy atoms in which higher 
order correlations are dominated by screening of the Coulomb interaction between 
core and valence electrons by other core electrons. In light atoms like 
germanium this effect does not dominate due to small number of electrons in the core.
Therefore, inclusion of screening, while other higher-order effects are
not included, does not improve the accuracy. 

The results of calculations are presented in Table~\ref{ge}.
The ground-state energies are given as energies to remove all valence
electrons from an atom or ion (in atomic units).
Corresponding experimental energies are sums of the ionization potentials 
of all relevant ions. For the convenience of comparison with Moore's 
tables~\cite{moore} we present energies of excited states relative to
the ground state in cm$^{-1}$. Column marked CI presents the results of
the standard configuration interaction method without $\hat \Sigma$.
Column $\hat \Sigma^{(2)}$ presents the results of calculations with
the effective Hamiltonian (\ref{heff}) in which $\hat \Sigma$ is
calculated in the second order of MBPT.

The results presented in Table~\ref{ge} show that inclusion of the 
core-valence correlations leads to systematic significant improvement
of the accuracy of calculations for all states of all ions and for
neutral germanium.

\begin{table}
\caption{\label{ge}Ground state removal energies (RE, a.u.) and excitation energies 
(cm$^{-1}$) of low states of Ge~IV to Ge~I.}
\begin{ruledtabular}
\begin{tabular}{llrrr}
\multicolumn{2}{c}{State}  &  CI &  $\hat \Sigma^{(2)}$ & Exp.~\cite{nist} \\
\hline
\multicolumn{5}{c}{Ge~IV} \\
\multicolumn{2}{l}{$4s_{1/2}$} \ RE &  -1.63631 & -1.68047 & -1.67993 \\
\multicolumn{2}{l}{$4p_{1/2}$} &   78746   &  81623   &  81315  \\  
\multicolumn{2}{l}{$4p_{1/2}$} &   81372   &  84470   &  84103  \\
\multicolumn{2}{l}{$4d_{1/2}$} &  183779   & 191142   & 190607  \\
\multicolumn{2}{l}{$4d_{1/2}$} &  184049   & 191424   & 190861  \\

\multicolumn{5}{c}{Ge~III} \\
$4s^2$   & $1S_0$ \ RE &   -2.85213 & -2.93114 &  -2.93765  \\
	     		 	          			      
$4s4p$   & $3P_0$ &      57762 &    61812 &     61734  \\   
         & $3P_1$ &      58490 &    62595 &     62500  \\    
         & $3P_2$ &      60030 &    64273 &     64144  \\   
	     		 	          			      
$4s4p$   & $1P_1$ &      90820 &    92238 &     91873  \\   

$4s4d$   & $1D_2$ &     137686 &   145305 &    144975  \\
	     		 	          			      
$4p^2$   & $3P_0$ &     142850 &   148023 &    147685 \\
         & $3P_1$ &     143721 &   148997 &    148640 \\
         & $3P_2$ &     145276 &   150765 &    150372 \\
	     		 	          			      
$4s5s$   & $3S_1$ &     152184 &   158630 &    158565  \\  

\multicolumn{5}{c}{Ge~II} \\

$4s^24p$ & $^2P^o_{1/2}$ \ RE & -3.42509 & -3.51488 &  -3.52322   \\
         & $^2P^o_{3/2}$ &     1623 &     1797 &      1767   \\
			      	   	     	 
$4s4p^2$ & $^4P_{1/2}$   &    47667 &    51512 &     51576   \\
         & $^4P_{3/2}$   &    48326 &    52241 &     52291   \\
         & $^4P_{5/2}$   &    49333 &    53342 &     53367   \\
			      	 			  
$4s^25s$ & $^2S_{1/2}$   &    61124 &    62870 &     62402  \\
			      	 				  
$4s4p^2$ & $^2D_{3/2}$   &    61750 &    65313 &     65015   \\
         & $^2D_{5/2}$   &    61930 &    65494 &     65184   \\
			      	  	  
$4s^25p$ & $^2P^o_{1/2}$ &    77370 &    79386 &     79006   \\
         & $^2P^o_{3/2}$ &    77710 &    79750 &     79366   \\
			      	 				  
$4s^24d$ & $^2D_{3/2}$   &    79270 &    81444 &     80836   \\
         & $^2D_{5/2}$   &    79439 &    81625 &     81012   \\
\multicolumn{5}{c}{Ge~I} \\
$4s^24p^2$ & $^3P_0$  \ RE & -3.70376  & -3.79871 &  -3.81352  \\
           & $^3P_1$     &      493  &      556 &       557  \\
           & $^3P_2$     &     1276  &     1423 &      1410  \\
			   	     
$4s^24p^2$ & $^1D_2$     &     7320  &     7591 &      7125  \\
			  							     
$4s^24p^2$ & $^1S_0$     &    17093  &    17541 &     16367  \\

$4s^24p5s$ & $^3P_0$     &    38969  &    38665 &     37452  \\
           & $^3P_1$     &    39272  &    38963 &     37702  \\
           & $^3P_2$     &    39024  &    40385 &     39118  \\
			  							     
$4s^24p5s$ & $^1P_1$     &    42010  &    41648 &     40020  \\

$4s^24p5p$ & $^1P_1$     &    45489  &    45503 &     45985  \\
			  							     
$4s^24p5p$ & $^3D_1$     &    46246  &   46199   &     46765  \\
           & $^3D_2$     &    46332  &   46275   &     46834  \\
           & $^3D_3$     &    47469  &   47620   &     48104  \\
\end{tabular}
\end{ruledtabular}
\end{table}
\subsection{Calculations for tin.}

Tin atom ($Z=50$) is very similar to the germanium atom. 
Its ground state configuration is $\dots 5s^25p^2$.
However, correlations and relativistic corrections are larger.
It has some implication on the calculation scheme. It turns out that
inclusion of the higher-order core-valence correlations does lead
to significant improvement of the results for all tin ions
and for the neutral atom. We include screening of Coulomb interaction 
and hole-particle interaction in all orders of the MBPT in the
calculation of $\hat \Sigma_1$. It is done exactly 
the same way as in our calculations for single-valence-electron atoms 
(see, e.g. \cite{Dzuba89}). The $\hat \Sigma_2$ operator is still 
calculated in the second order of the MBPT. 

The results are presented in Table~\ref{sn}. 
There is one more column in the table compared to Table~\ref{ge}.
It is marked $\hat \Sigma^{(\infty)}$ and presents the results of calculations
with all-order $\hat \Sigma_1$. Again, it easy to see that moving from less
sophisticated to more sophisticated  approximations (with no~$\hat \Sigma$;
with $\hat \Sigma^{(2)}$; with $\hat \Sigma^{(\infty)}$) leads to systematic
significant improvement of the accuracy of the results.

\begin{table}
\caption{\label{sn}Ground state removal energies (RE, a.u.) and excitation energies 
(cm$^{-1}$) of low states of Sn~IV to Sn~I.}
\begin{ruledtabular}
\begin{tabular}{llrrrr}
\multicolumn{2}{c}{State}  &  CI &  $\hat \Sigma^{(2)}$ &
   $\hat \Sigma^{(\infty)}$ & Exp.~\cite{moore} \\
\hline
\multicolumn{6}{c}{Sn~IV} \\
$4d^{10}5s$ & $^2S_{1/2}$ \ RE & -1.43894 & -1.51228 & -1.49776 & -1.49699 \\
$4d^{10}5p$ & $^2P_{1/2}$        &    66323 &    70709 &    69727 &    69564 \\
            & $^2P_{3/2}$        &    72291 &    77409 &    76264 &    76072 \\
$4d^{10}5d$ & $^2D_{3/2}$        &   156481 &   168074 &   165406 &   165305 \\
            & $^2D_{5/2}$        &   157180 &   168847 &   166183 &   165411 \\
\multicolumn{6}{c}{Sn~III} \\
$5s^2$ & $^1S_0$ \ RE 
                   & -2.51142 & -2.64097 &  -2.61447 & -2.61794 \\ 
					   	      		 
$5s5p$ & $^3P^o_0$ &    47961 &    54914 &     54001 &   53548 \\
       & $^3P^o_1$ &    49548 &    56582 &     55631 &   55196 \\
       & $^3P^o_2$ &    53207 &    60734 &     59670 &   59229 \\
					   	      		 
       & $^1P^o_1$ &    78801 &    80163 &     79019 &   79911 \\
					   	      		 
$5p^2$ & $^3P_0$   &   121290 &   128814 &    126873 &  127309 \\
       & $^3P_1$   &   123690 &   131743 &    129709 &  130120 \\
       & $^3P_2$   &   118412 &   136470 &    134275 &  134567 \\
					   	      		 
       & $^1D_2$   &   127379 &   130638 &    128478 &  128205 \\
					   	      		 
$5s6s$ & $^3S_1$   &   130986 &   141420 &    139341 &  139638 \\
					   	      		 
$5s5d$ & $^3D_1$   &   132760 &   142898 &    140463 &  141322 \\
       & $^3D_2$   &   132946 &   143107 &    140671 &  141526 \\
       & $^3D_3$   &   133222 &   143423 &    140987 &  141838 \\
					   	      		 
$5s6s$ & $^1S_0$   &   135453 &   145105 &    143064 &  143591 \\
						      		 
$5s5d$ & $^1D_2$   &   148378 &   155394 &    153063 &  154116 \\

\multicolumn{6}{c}{Sn~II} \\

$5s^25p$ & $^2P^o_{1/2}$ \ RE
                         &   -3.03218 & -3.17791 & -3.14624  & -3.15567 \\
         & $^2P^o_{3/2}$ &       3776 &     4352 &     4222  &     4251 \\
			   	   
$5s5p^2$ & $^4P_{1/2}$   &      40839 &    47579 &    46661  &    46464 \\
         & $^4P_{3/2}$   &      42512 &    49537 &    48556  &    48368 \\
         & $^4P_{5/2}$   &      44720 &    51958 &    50915  &    50730 \\
			   	   
$5s^26s$ & $^2S_{1/2}$   &      54896 &    57545 &    56707  &    56886 \\
			   	   
$5s5p^2$ & $^2D_{3/2}$   &      54142 &    59969 &    58806  &    58844 \\
         & $^2D_{5/2}$   &      54731 &    60599 &    59419  &    59463 \\
			   	   
$5s^25d$ & $^2D_{3/2}$   &      69220 &    72247 &    71140  &    71406 \\
         & $^2D_{5/2}$   &      69776 &    72929 &    71804  &    72048 \\
			   	   
$5s^26p$ & $^2P^o_{1/2}$ &      69006 &    72131 &    71182  &    71494 \\
         & $^2P^o_{3/2}$ &      69825 &    73025 &    72061  &    72377 \\
\multicolumn{6}{c}{Sn~I} \\
$5s^25p^2$ &  $^3P_0$ \ 
                  RE &  -3.28899  &  -3.44213  & -3.407850  &  -3.425548  \\
           &  $^3P_1$  &      1411  &      1681  &      1623  &       1692  \\
           &  $^3P_2$  &      3049  &      3539  &      3428  &       3428  \\

       	   &  $^1D_2$  &      8359  &      9079  &      8891  &       8613  \\

           &  $^1S_0$  &     17328  &     18217  &     17977  &      17163  \\

$5s^25p6s$ &  $^3P_0$  &     35381  &     35722  &     35251  &      34641  \\
           &  $^3P_1$  &     35764  &     36050  &     35577  &      34914  \\
           &  $^3P_2$  &     38988  &     39848  &     39252  &      38629  \\

           &  $^1P_1$  &     40080  &     40655  &     40063  &      39257  \\

$5s5p^3$   &  $^5S_2$  &     34720  &     40529  &     39725  &      39626  \\

$5s^25p6p$ &  $^3P_0$  &     42805  &     44164  &     43578  &      43430  \\
           &  $^3P_1$  &     41361  &     42785  &     42200  &      42342  \\
           &  $^3P_2$  &     45804  &     47712  &     47008  &      47235  \\

$5s^25p6p$ &  $^3D_1$  &     42356  &     43768  &     43178  &      43369  \\
           &  $^3D_2$  &     42447  &     43861  &     43267  &      43239  \\
           &  $^3D_3$  &     45543  &     47511  &     46796  &      47007  \\
\end{tabular}
\end{ruledtabular}
\end{table}
\subsection{Calculations for lead}

The case of lead ($Z=82$) is the most difficult of the calculations. 
Correlations are strong and relativistic effects are large too.
Strong $L-S$ interaction leads to intersection of the fine-structure 
multiplets. Also, states of the same total momentum $J$ are
strongly mixed regardless of the values of $L$ and $S$ assigned to them.
The breaking of the $L-S$ scheme can be easily seen e.g. by comparing
experimental values of the Land\'{e} $g$-factors with the non-relativistic
values.

We have done one more step for lead to further improve the accuracy
of calculations as compared to the scheme used for tin. We have
introduced the scaling factors before $\hat \Sigma_1$ to fit the energies
of Pb~IV. These energies are found by solving Hartree-Fock-like
equations for the states of external electron of Pb~IV in the $V^{N-4}$
potential of the atomic core

\begin{equation}\label{Brueck}
  (\hat H_0 + \hat \Sigma_1 - \epsilon_n)\psi_n = 0.
\end{equation}

Here $\hat H_0$ is the Hartree-Fock Hamiltonian. $\hat \Sigma_1$ is 
the all-order correlation potential operator similar to what is used for tin.
Inclusion of $\hat \Sigma_1$ takes into account the effect of the core-valence 
correlations on both the energies ($\epsilon_n$) and the wave functions
($\psi_n$) of the valence states producing the so-called Brueckner orbitals.
The difference between Brueckner and experimental energies of the 
$4s$, $4p$ and $4d$ states of Pb~IV are on the level of 0.2 - 0.4\%
(for removal energies). 
To further improve the energies we replace $\hat \Sigma_1$
by $f \hat \Sigma_1$ with rescaling factor $f$ chosen to fit the energies
exactly. Then the same rescaled operator $f \hat \Sigma_1$ is used for
the Pb~III and Pb~II ions and for the Pb~I. It turns out that only small rescaling 
is needed. Maximum deviation of the rescaling factor from unity is  10\%:
$ f(4s)=0.935, \ f(4p_{1/2}) = 1.084, \ f(4p_{3/2}) = 1.1, \
f(4d_{3/2}) = 1.07, \ f(4d_{5/2}) = 1.07. $

The results of the calculations are presented in Table~\ref{pb}. 
Again, inclusion of core-valence correlations lead to significant
improvement of the accuracy of the results in all cases.
However, comparison between different ways of treating core-valence
correlations reveal a more complicated picture compared to what we
have for tin. When we move from the second-order correlation operator
$\hat \Sigma^{(2)}$ to the all-order  $\hat \Sigma^{(\infty)}$
and then to the rescaled  $f\hat \Sigma^{(\infty)}$ the improvement
in accuracy is apparent for the removal energies. It is again
systematic and significant, bringing results for all states of
all ions and neutral lead to better agreement with experiment.
This is not always the case for the energy intervals. When
a more accurate treatment of core-valence correlation is
introduced two energy levels way move cowards experimental values at
slightly different rate so that the interval between them does not
improve. In Table~\ref{pb} we present removal energies only for the 
ground states of Pb~IV, Pb~III, Pb~II and Pb~I. Energies of excited
states are given with respect to the ground state. It is easy to 
see that energy intervals between ground and excited states
calculated with second-order $\hat \Sigma$ are often in better
agreement with experiment than the results with the all-order
$\hat \Sigma$. In general, the results are not as good as for tin.
The reason for this is larger overlap between valence and core states.
Relativistic effects cause stronger binding of the $6s$ and $6p$
electrons of Pb compared to binding of the $5s$ and $5p$ electrons
of Sn. This means that overlap between valence and core states
is also larger for lead than for tin leading to larger effect of 
removal of valence electrons on atomic core.

It is instructive to compare our results with the results of
recent calculations by Safronova {\em et al}~\cite{Safronova}
(see Table~\ref{pb}).
Energy levels of Pb~II were calculated by Safronova {\em et al}
with the use of the coupled-cluster (CC) approach and the third-order MBPT.
The Pb~II ion was treated as an ion with one external electron above
closed shells. Therefore only energies of states in  which the $6s$
subshell remained closed were calculated. The agreement with
experiment for these states is slightly better than for our results 
with $\hat \Sigma^{(\infty)}$. The reason for this is better treatment
of the interaction between core and valence electrons.
The $6s$ electrons were included in the initial Hartree-Fock
procedure. Also, interaction between the $6p$ electron and the core
is included in the CC approach in all-orders of the MBPT.

This doesn't mean that the $V^{N-4}$ approximation is not good for 
lead. First, as can be seen from Table~\ref{pb}, inclusion of core-valence
correlation does lead to systematic significant improvement of the accuracy
and final results are very close to the experiment. Second, the fact that
inclusion of the higher order core-valence correlations doesn't always
lead to improvement of energy intervals doesn't mean that the $V^{N-4}$
approximation is not good. It rather means that not all dominating
higher-order diagrams are included into $\hat \Sigma^{(\infty)}$. The situation
is very similar to what takes place for single-valence-electron atoms.
The technique developed by us for alkali atoms~\cite{Dzuba89}
doesn't work very well for atoms like thallium where interaction between
valence electron and the core is important. 
Here CC+MBPT approach gives better results~\cite{Safronova} which
may mean that the combination of the CC approach with the CI method
is a better option for atoms like lead. This approach was recently
considered by Kozlov~\cite{Kozlov04} and Johnson~\cite{Johnson}.
However, no calculations for real atoms have been done so far.

\begin{table*}
\caption{\label{pb}Ground state removal energies (RE, a.u.) and excitation energies 
(cm$^{-1}$) of low states of Pb~IV to Pb~I.}
\begin{ruledtabular}
\begin{tabular}{llrrrrrr}
\multicolumn{2}{c}{State}  &  CI &  $\hat \Sigma^{(2)}$ &
 $\hat \Sigma^{(\infty)}$ & $f\hat \Sigma^{(\infty)}$
& Ref.~\cite{Safronova} & Exp.~\cite{moore} \\
\hline
\multicolumn{8}{c}{Pb~IV} \\
$5d^{10}6s$ & $^2S_{1/2}$ \ RE & -1.48374 & -1.57689 & -1.56035 & -1.55529 & & -1.55531 \\
$5d^{10}6p$ & $^2P_{1/2}$        &    72857 &    78055 &    78239 &    76144 & &   76158 \\
            & $^2P_{3/2}$        &    92301 &    99817 &    99388 &    97276 & &   97219 \\
$5d^{10}6d$ & $^2D_{3/2}$        &   173446 &   188501 &   185992 &   184570 & &  184559 \\
            & $^2D_{5/2}$        &   175485 &   190789 &   188254 &   186848 & &  186817 \\
\multicolumn{8}{c}{Pb~III} \\
$6s^2$ & $^1S_0$ \ RE & -2.58923 & -2.76503  & -2.73356 & -2.72421 &  & -2.72853 \\
								       	
$6s6p$ & $^3P_0$ &    52866  &       62881  &      62947 &     61045  &  &    60397 \\
       & $^3P_1$ &   57184   &      66767   &     66751  &    64851   &  &    64391 \\
       & $^3P_2$ &   70223   &      82032   &     81477  &    79577   &  &    78985 \\
								       	
       & $^1P_1$ &   91945   &      96556   &     95876  &    94071   &  &    95340 \\
								       	
$6p^2$ & $^3P_0$ &  135286   &     145385   &    145400  &   141555   &  &   142551 \\
$6s7s$ & $^3S_1$ &  137664   &     153445   &    150863  &   150038   &  &   150084 \\
$6s6d$ & $^1D_2$ &  138279   &     156137   &    154498  &   152079   &  &   151885 \\
$6s7s$ & $^1S_0$ &  142139   &     156815   &    154219  &   153407   &  &   153783 \\
\multicolumn{8}{c}{Pb~II} \\
$6s^26p$ & $^2P^o_{1/2}$ \ RE &  -3.11363 & -3.31759 & -3.27430 & -3.26897 & & -3.28141 \\
         & $^2P^o_{3/2}$ &     12390 &    14447 &    13858 &    13896 & 14137 &    14081 \\

$6s6p^2$ & $^4P_{1/2}$   &     50298 &    59934 &    59934 &    58052 &       &    57911 \\
         & $^4P_{3/2}$   &     57209 &    68501 &    67633 &    66221 &       &    66124 \\
         & $^4P_{5/2}$   &     61484 &    75957 &    74856 &    73749 &       &    73905 \\

$6s^27s$ & $^2S_{1/2}$   &     55451 &    60525 &    58170 &    59203 & 58967 &    59448 \\

$6s^26d$ & $^2D_{5/2}$   &     66823 &    71130 &    69314 &    69256 & 70229 &    68964 \\
         & $^2D_{3/2}$   &     63732 &    70711 &    68916 &    69001 & 69686 &    69740 \\

$6s^27p$ & $^2P^o_{1/2}$ &     69961 &    75342 &    73140 &    73878 & 74256 &    74459 \\
         & $^2P^o_{3/2}$ &     72572 &    78180 &    75935 &    76666 & 77069 &    77272 \\

$6s6p^2$ & $^2D_{3/2}$   &     77272 &    85538 &    84523 &    83196 &       &    83083 \\
         & $^2D_{5/2}$   &     81630 &    91291 &    89614 &    88800 &       &    88972 \\
\multicolumn{8}{c}{Pb~I} \\
$6s^26p^2$ &  $^3P_0$ \ RE & -3.36433  &  -3.58255  & -3.53174  & -3.52974  & &-3.55398 \\
           &  $^3P_1$  &     6388  &      7736  &     7305  &     7353  & &    7819 \\
           &  $^3P_2$  &       9199  &     10795  &    10277  &    10423  & &   10650 \\

       	   &  $^1D_2$  &      18578  &     21793  &    20780  &    20979  & &   21458 \\
						      	 	  	 
           &  $^1S_0$  &      26998  &     30355  &    29185  &    29412  & &   29467 \\

$6s^26p7p$ &  $^3P_0$  &      33413  &     35239  &    33679  &    34517  & &   34960 \\
           &  $^3P_1$  &      33871  &     35610  &    34056  &    34887  & &   35287 \\

$6s^26p7p$ &  $^3P_1$  &      40029  &     42987  &    41405  &    42061  & &   42919 \\
           &  $^3P_0$  &      41612  &     44441  &    42882  &    43525  & &   44401 \\

$6s^26p7p$ &  $^3D_1$  &      41740  &     44714  &    43129  &    43773  & &   44675 \\
           &  $^3D_2$  &      41886  &     44868  &    43281  &    43958  & &   44809 \\
\end{tabular}
\end{ruledtabular}
\end{table*}

\section{Conclusion}

It has been demonstrated that the $V^{N-4}$ approximation works very 
well for the four-valence-electrons atoms like germanium, tin and lead
as well as for their single, double and triple ionized ions. The use
of the $V^{N-4}$ approximation makes it easy to include core-valence
correlations beyond the second order of the MBPT. Inclusion of the
core-valence correlations leads to significant improvement of the results 
in all cases. In general, the $V^{N-M}$ approximation ($M$ is the number
of valence electrons) is a good approximation if the overlap between core
and valence states is small. The best case is the alkaline-earth atoms
where the $V^{N-2}$ approximation must produce excellent results.
In contrast, the $V^{N-M}$ approximation is not applicable at all
to atoms with open $d$ or/and  $f$ shells unless uppermost core
$s$ and $p$ states are also treated as valence states.
It should work more or less
well for most of the atoms/ions with $s$ and/or $p$ valence electrons.
In cases of relatively large overlap between core and valence states
good results can still be achieved if accurate treatment of the 
interaction between core and valence electrons is included 
perturbatively into the calculation of the core-valence correlations.

\section{Acknowledgments}

The author is grateful to J. S. M. Ginges and V. V. Flambaum for 
useful discussions.



\begin{thebibliography}{20}
\bibitem{comment} Strictly speaking, different values of $q$ for
different configurations would unavoidably lead to level crossing
because $q$ is a slope of the $E(\alpha^2)$ graph. When we say ``there
is no level crossing'' we mean that level crossing takes place far
away from $\alpha=\alpha_0$.
\end{thebibliography}
\end{document}